\documentclass[12pt,preprint,prc,aps,showpacs,showkeys,groupedaddress,floatfix]{revtex4}
\usepackage{epsfig}
\usepackage{dcolumn}
\usepackage{bm}
\usepackage{graphics}
\usepackage{graphicx}
\usepackage{amssymb}
\usepackage{amsmath}
\begin{document}

\title{Analytical solutions of the Bohr Hamiltonian with the Morse potential}

\author{I. Boztosun\S, D. Bonatsos\dag, and I. Inci\S}
\affiliation{\S Department of Physics, Erciyes University, Kayseri, Turkey}
\affiliation{\dag Institute of Nuclear Physics, N.C.S.R. ``Demokritos",
GR-15310 Aghia Paraskevi, Attiki, Greece}

\date{\today}

\begin{abstract}

Analytical solutions of the Bohr Hamiltonian are obtained in the
$\gamma$-unstable case, as well as in an exactly separable
rotational case with $\gamma\approx 0$, called the exactly separable Morse
(ES-M) solution. Closed expressions
for the energy eigenvalues are obtained through the Asymptotic
Iteration Method (AIM), the effectiveness of which is demonstrated
by solving the relevant Bohr equations for the Davidson and Kratzer
potentials. All medium mass and heavy nuclei with known $\beta_1$
and $\gamma_1$ bandheads have been fitted by using the two-parameter $\gamma$-unstable
solution for transitional nuclei and the three-parameter ES-M for rotational ones.
It is shown that  bandheads and energy spacings within the bands
are well reproduced for more than 50 nuclei in each case.

\end{abstract}


\maketitle

\section{Introduction}

The recent introduction of the critical point symmetries E(5) \cite{IacE5} and X(5) \cite{IacX5},
which describe shape phase transitions between vibrational and $\gamma$-unstable/prolate
deformed rotational nuclei respectively, has stirred much interest in special solutions
of the Bohr Hamiltonian \cite{Bohr}, describing collective nuclear properties in terms of the
collective variables $\beta$ and $\gamma$.
Such solutions can describe nuclei in the whole region
between different limiting symmetries, while critical point symmetries are appropriate for
describing nuclei only at or near the critical point, in good agreement with experiment
\cite{CZE5,Zamfir,CZX5,Kruecken,JPGreview}.

Shape phase transitions in nuclear structure have been first discovered \cite{Deans}
in the classical analog \cite{GK,DSI} of the Interacting Boson Model \cite{IA},
which describes collective nuclei in terms of collective bosons of angular momentum zero
($s$-bosons) and two ($d$-bosons) in the framework of a U(6) overall symmetry,
possessing U(5) (vibrational), SU(3) (prolate deformed rotational), and O(6)
($\gamma$-unstable) limiting symmetries. To visualize these limiting symmetries and the
transitions between them it is useful to place them at the corners of a symmetry triangle
\cite{triangle}. A similar triangle for the collective model has been introduced \cite{ctriangle}.
In the IBM framework it has been found that a first order phase transition occurs between U(5) and SU(3),
while a second order phase transition occurs between U(5) and O(6) \cite{Deans}.
Within the collective model, X(5) corresponds to the first case and E(5) to the second.

It has been known for a long time \cite{Wilets} that simple special solutions of the Bohr
Hamiltonian, resulting from exact separation of variables in the relevant Schr\"odinger equation,
can be obtained in the $\gamma$-unstable case, in which the potential depends
only on $\beta$, as well as in the case in which the potential can be written in the separable form
\begin{equation}\label{eq:ES}
u(\beta,\gamma) = u(\beta) + {u(\gamma)\over \beta^2},
\end{equation}
in the special cases of $\gamma \approx 0$ or $\gamma\approx \pi/6$ \cite{MtV}. An approximate separation
of variables has also been attempted for potentials of the form
\begin{equation}\label{eq:NES}
u(\beta,\gamma)=u(\beta)+u(\gamma)
\end{equation}
in the cases of $\gamma \approx 0$ \cite{IacX5} or $\gamma \approx \pi/6$ \cite{Z5}.
A brief summary of existing solutions is listed here.

1) The E(5) critical point symmetry \cite{IacE5} is a $\gamma$-unstable solution, using
as $u(\beta)$ an infinite-well potential
starting from $\beta=0$. Displacing the well from $\beta=0$ leads to the
O(5)-Confined Beta Soft [O(5)-CBS] model \cite{O5CBS}.
$\gamma$-unstable solutions have also been given for the Coulomb \cite{Fortunato2} and
Kratzer \cite{Fortunato2} potentials, $\beta^{2n}$ potentials ($n=1$, 2, 3, 4) \cite{Ariasb4,BonE5}
(labelled as E(5)-$\beta^{2n}$),
as well as for the Davidson potential \cite{Dav,Elliott,Bahri,varPLB}
\begin{equation}\label{Davidsonpotential}
u(\beta)=\beta^2+{\beta_0^4\over\beta^2},
\end{equation}
where $\beta_0$ is the position of the minimum of the potential.
A solution using a well of finite depth has also been given \cite{finite}.

2) The X(5) critical point symmetry \cite{IacX5}  is an approximate solution, using a potential of
the form of Eq. (\ref{eq:NES}) with $\gamma \approx 0$ [achieved by using as $u(\gamma)$ a harmonic
oscillator potential with minimum at $\gamma=0$]
and an infinite-well potential
starting from $\beta=0$ as $u(\beta)$. Displacing the well from $\beta=0$ leads to the
Confined Beta Soft (CBS) model \cite{CBS}. Similar solutions
have been obtained for $\beta^{2n}$ potentials ($n=1$, 2, 3, 4) \cite{BonX5} (labelled as
X(5)-$\beta^{2n}$),
as well as for the Davidson potential \cite{varPLB}.

3) Z(5) \cite{Z5} is an approximate solution, using a potential of
the form of Eq. (\ref{eq:NES}) with $\gamma \approx \pi/6$.

4) Exactly separable solutions using a potential of the form of Eq. (\ref{eq:ES})
and $\gamma \approx 0$ have been obtained for the  Coulomb \cite{Fortunato1} and
Kratzer \cite{Fortunato1} potentials, the infinite-well potential [labelled as
ES-X(5)] \cite{ESX5}, the harmonic oscillator potential (labelled as ES-X(5)-$\beta^2$) \cite{ESX5},
as well as for the Davidson potential \cite{ESD} (labelled as ES-D).

5) Exactly separable solutions, using a potential of the form of Eq. (\ref{eq:ES})
and $\gamma \approx \pi/6$, have been obtained for the Coulomb \cite{Fort70,Fort74},
Kratzer \cite{Fort70,Fort74}, and Davidson \cite{Fort70,Fort74} potentials.

6) In addition to the special analytical solutions mentioned above, a powerful method
for solving the Bohr Hamiltonian numerically has been developed recently \cite{Rowe735},
evolving into an algebraic collective model \cite{Rowe753}. The relations between the
algebraic collective model and the different limiting symmetries of the Interacting Boson
Model \cite{IA} have been studied in Refs. \cite{Thiamova1,Thiamova2}. Using this numerical
method, the Bohr Hamiltonian has been solved \cite{Caprio} for the same potentials used in X(5), but avoiding
the approximate separation of variables, resulting in evidence for strong
beta--gamma mixing.

Chains of models mentioned above cover regions between different limiting symmetries.
For example, the chain E(5)-$\beta^{2n}$ ($n=1$, 2, 3 ,4), E(5), O(5)-CBS spans
the region between the vibrational [U(5)] and $\gamma$-unstable [O(6)] limits,
while the chain X(5)-$\beta^{2n}$, X(5), CBS spans the region between the vibrational
and prolate deformed rotational [SU(3)] limits. Their predictions can therefore be tested against
a large body of experimental data \cite{Nature}.

The potentials mentioned above (infinite well, harmonic oscillator, Coulomb, Kratzer, Davidson)
are known to be exactly soluble for all values of angular momentum $L$. In the present work, we introduce
special solutions for the Morse potential \cite{Morse},
\begin{equation}\label{Morsepot}
u(\beta)=e^{-2a(\beta-\beta_e)}-2e^{-a(\beta-\beta_e)},
\end{equation}
which is known \cite{Flugge,SUSYQM} to be exactly soluble only for
$L=0$. Analytical expressions for the spectra for any $L$ are
obtained by solving the relevant differential equation through the
Asymptotic Iteration Method (AIM) \cite{hakanaim1,hakanaim2}, after
applying the Pekeris approximation \cite{Pekeris}. Solutions for the
$\gamma$-unstable case and the exactly separable rotational case
with $\gamma \approx 0$ (to be called ES-M) are obtained. In order
to demonstrate the effectiveness of AIM, we first apply it to the
Davidson and Kratzer potentials in the same cases
($\gamma$-unstable, exactly separable rotational with $\gamma\approx
0$), recovering the above-mentioned solutions which can be obtained
in terms of special functions.

A few advantages of the present approach are listed here.

1) In X(5) and related models, using potentials of the form of Eq.
(\ref{eq:NES}), the ground state and beta bands depend only on the
parameters of the $\beta$ potential, while the gamma bands depend
also on an additional parameter introduced by the $\gamma$ potential
[usually the stiffnes of the harmonic oscillator used as
$u(\gamma)$]. When exactly separable potentials of the form of Eq.
(\ref{eq:ES}) are used, all bands (ground state, beta, gamma) depend
on all parameters. Thus, all bands are treated on an equal footing,
as in the case of the ES-D solution \cite{ESD}.

2) A well known problem of X(5) and related solutions is the
overprediction of the energy spacings within the beta band by almost
a factor of two \cite{CZX5,Kruecken,JPGreview}. It is known that
this problem can be avoided by replacing the infinite-well potential
of X(5) by a potential with sloped walls \cite{sloped}. The present
solution avoids this problem, since the right branch of the Morse
potential imitates the sloped wall.

In order to test the applicability of the Morse potential in the description of nuclear spectra,
we have fitted all nuclei with mass $A\geq 100$ and $R_{4/2}=E(4)/E(2)<2.6$ for which at least
the $\beta_1$ and $\gamma_1$ bandheads are known \cite{NDS}, using the $\gamma$-unstable solution of the Morse
potential, which involves two free parameters ($\beta_e$, $a$). We have also fitted all nuclei
with mass $A\geq 150$ and $R_{4/2}=E(4)/E(2)>2.9$ for which at least
the $\beta_1$ and $\gamma_1$ bandheads are known \cite{NDS}, using the exactly separable rotational solution of the Morse
potential with $\gamma \approx 0$ (ES-M), which involves three free parameters (the Morse parameters $\beta_e$ and $a$,
as well as the stiffness $c$ of the $\gamma$ potential, for which a harmonic oscillator is used).
A comparison of the latter to the fits provided by the Davidson potential in the exactly separable $\gamma\approx 0$ case
\cite{ESD} (ES-D), which contains two free parameters ($\beta_0$, $c$) instead of three, shows that the extra
parameter extends the region of applicability of the model in the same nuclei to higher
angular momenta, largely improving the quality of the fits.

In Section II of the present work, the Asymptotic Iteration Method
(AIM) is briefly reviewed. The method is then applied to the exactly
separable rotational $\gamma\approx 0$ case for the Davidson,
Kratzer, and Morse potentials in Section III, and to the
$\gamma$-unstable case of the same potentials in Section IV. Fits to
experimental data are presented in Section V, while Section VI
contains discussion of the present results and plans for further
work.

\section{Overview of the Asymptotic Iteration Method}
\label{aim}

The Asymptotic Iteration Method (AIM) has been
proposed \cite{hakanaim1,hakanaim2} and applied
\cite{karakoc,bayrakJPA,boztosun,Soylu, bayrak, bayrak1} to the solution of
second-order differential equations of the form
\begin{equation}\label{differential}
  y''=\lambda_{0}(x)y'+s_{0}(x)y,
\end{equation}
where $\lambda_{0}(x)\neq 0$ and the prime denotes the derivative
with respect to $x$. The functions, $s_{0}(x)$ and $\lambda_{0}(x)$,
must be sufficiently differentiable. Eq.
(\ref{differential}) has a general solution \cite{hakanaim1}
\begin{equation}\label{generalsolution}
  y(x)=exp \left( - \int^{x} \alpha(x_{1}) dx_{1}\right ) \left [C_{2}+C_{1}
  \int^{x}exp  \left( \int^{x_{1}} [\lambda_{0}(x_{2})+2\alpha(x_{2})] dx_{2} \right ) dx_{1} \right
  ]
\end{equation}
for sufficiently large $k$, $k>0$, if
\begin{equation}\label{termination}
\frac{s_{k}(x)}{\lambda_{k}(x)}=\frac{s_{k-1}(x)}{\lambda_{k-1}(x)}=\alpha(x),
\end{equation}
where
\begin{eqnarray}\label{iteration}
  \lambda_{k}(x) & = &
  \lambda_{k-1}'(x)+s_{k-1}(x)+\lambda_{0}(x)\lambda_{k-1}(x), \quad
  \nonumber \\
s_{k}(x) & = & s_{k-1}'(x)+s_{0}(x)\lambda_{k-1}(x), \quad \quad
\quad \quad k=1,2,3,\ldots
\end{eqnarray}
For a given potential, the radial Schr\"{o}dinger equation is
converted to the form of Eq. (\ref{differential}). Then, s$_{0}(x)$
and $\lambda_{0}(x)$ are determined, and the functions s$_{k}(x)$ and
$\lambda_{k}(x)$ are calculated by the recurrence
relations of Eq. ~(\ref{iteration}).

The termination condition of the method, given in Eq. (\ref{termination}),
can be arranged as
\begin{equation}\label{quantization}
  \Delta_{k}(x)=\lambda_{k}(x)s_{k-1}(x)-\lambda_{k-1}(x)s_{k}(x)=0, \quad \quad
k=1,2,3,\ldots
\end{equation}
Then, the energy eigenvalues are obtained from the roots of Eq.
(\ref{quantization}) if the problem is exactly solvable. If not, for
a specific principal quantum number $n$, we choose a suitable $x_0$
point, generally determined as the maximum value of the asymptotic
wave function or the minimum value of the potential
\cite{hakanaim1,boztosun}, and the approximate energy eigenvalues
are obtained from the roots of this equation for sufficiently large
values of $k$ by iteration.

The corresponding eigenfunctions can be derived from the following
wave function generator for exactly solvable potentials
\begin{equation}\label{generator}
y_n (x) = C_2 \exp \left( { - \int\limits^x
{\frac{s_{n}(x_{1})}{\lambda_{n}(x_{1})}dx_{1}} } \right),
\end{equation}
where $n$ represents the radial quantum number.

Recently, Boztosun and Karakoc \cite{boztosun1} have further improved the method for the exactly solvable problems by rewriting the second-order differential equation of Eq. (\ref{differential}) in the form
\begin{equation}\label{newdifferential}
y''=-{\tau(x)\over\sigma(x)}y'-{\Omega_n\over\sigma(x)}y.
\end{equation}
By comparison with Eq. (\ref{differential}), $\tau(x)$,
$\sigma(x)$, and $\Omega_n$ can be found to be
\begin{equation}
-{\tau(x)\over\sigma(x)}=\lambda_0(x), \qquad -{\Omega_n\over\sigma(x)}=s_0(x),
\end{equation}
where $\Omega_n$ is a constant which comprises the eigenvalue. Then, the energy eigenvalues are obtained from
\begin{equation}\label{gamman}
\Omega_n = -n\sigma^{'}(x)-{n(n-1)\over 2}\sigma^{''}(x).
\end{equation}
The applicability of this new solution is demonstrated for the Morse
potential in Appendices A3 and A6.

\section{Exactly separable solutions for $\gamma \approx 0$}
\label{aimapp}

The original collective Bohr Hamiltonian \cite{Bohr} is
\begin{equation}\label{Bhamiltonian}
H = -{\hbar^2 \over 2B} \left[ {1\over \beta^4} {\partial \over
\partial \beta} \beta^4 {\partial \over \partial \beta} + {1\over
\beta^2 \sin 3\gamma} {\partial \over \partial \gamma} \sin 3 \gamma
{\partial \over
\partial \gamma} - {1\over 4 \beta^2} \sum_{k=1,2,3} {Q_k^2 \over \sin^2
\left(\gamma - {2\over 3} \pi k\right) } \right] +V(\beta,\gamma),
\end{equation}
where $\beta$ and $\gamma$ are the usual collective
coordinates which define the shape of the nuclear surface. $Q_k$
($k$=1, 2, 3) represents the angular momentum components in the
intrinsic frame, and $B$ is the mass parameter. Reduced energies and
reduced potentials are defined as $\epsilon = 2B E /\hbar^2$, $v= 2B
V /\hbar^2$ respectively \cite{IacE5}.
If the potential has a minimum around
$\gamma=0$, the angular momentum term in Eq. (\ref{Bhamiltonian}) can
be written  \cite{IacX5} as
\begin{equation}\label{Sum}
\sum _{k=1}^{3} {Q_k^2 \over \sin^2 \left( \gamma -{2\pi \over 3}
k\right)} \approx {4\over 3} (Q_1^2+Q_2^2+Q_3^2) +Q_3^2 \left(
{1\over \sin^2\gamma} -{4\over 3}\right).
\end{equation}

Exact separation of variables can be achieved
\cite{Wilets,Fort70,Fort74,ESD} for potentials of
the form $u(\beta,\gamma)=u(\beta)+u(\gamma)/\beta^2$, given
in Eq. (\ref{eq:ES}).
We then assume wavefunctions of the form
\begin{equation}\label{Wavefunction}
\psi(\beta,\gamma,\theta_j)=\xi_L(\beta)\Gamma_K(\gamma){\cal
D}_{M,K}^L(\theta_j),
\end{equation}
where $\theta_j$ ($j=1$, 2, 3) are the Euler
angles, ${\cal D}(\theta_j)$ represents Wigner functions of these
angles, $L$ stands for the eigenvalues of the angular momentum, while $M$ and $K$ are
the eigenvalues of the projections of the angular momentum on the
laboratory-fixed $z$-axis and the body-fixed $z'$-axis respectively.
The Schr\"odinger equation is thus separated, as in Refs.
\cite{Wilets,Fort70,Fort74,ESD}, into a  ``radial" part
(depending on $\beta$) and a $\gamma$ part.

\subsection{Davidson potential}

Solving through AIM, the $\beta$ equation for the Davidson potential
\cite{Dav} of Eq. (\ref{Davidsonpotential}),
$u(\beta)=\beta^2+{\beta_0^4/ \beta^2}$,
where $\beta_0$ is the position of the minimum of the potential,
we get the energy eigenvalues
\begin{equation}\label{DX5Eigenvalue}
\epsilon_{n,L}=2n+1+{\left[{9\over4}+{L(L+1)\over3}+\lambda+\beta_0^4\right]}^{1/2},
\end{equation}
where $\lambda$ is a term coming from the exact separation of variables, determined
from the $\gamma$ equation.

The $\gamma$ equation has been solved \cite{IacX5,ESD} for  a potential
\begin{equation}\label{eq:gamma}
u(\gamma)={(3c)^2\gamma^2},
\end{equation}
 leading to
\begin{equation}\label{eq:lambda}
 \lambda = \epsilon_\gamma -{K^2 \over 3}, \qquad \epsilon_\gamma = (3C) (n_\gamma+1),
\end{equation}
where $C=2c$. The final result for the energy eigenvalues coincides
with the results of Ref. \cite{ESD}.
The details of the AIM calculation are given in Appendix A1.

\emph{\bf Special case 1:}
If we take $\beta_0=0$, the Davidson potential becomes the potential of a
harmonic oscillator. Then, the expression for the energy eigenvalues coincides then
with the one obtained for the ES-X(5)-$\beta^2$ model of Ref. \cite{ESX5}.

\emph{\bf Special case 2:} Taking $c=0$ and $K=0$ (\emph{i.e.}
$\lambda=0$) in Eq. (\ref{DX5Eigenvalue}), we obtain for the ground
state and beta bands energy eigenvalues of the X(5)-Davidson
solution of Ref. \cite{varPLB}.

\subsection{Kratzer potential}

We again use an exactly separable potential of the form given in Eq.
(\ref{eq:ES}), with a Kratzer potential \cite{Fortunato1,Fortunato2}
\begin{equation}\label{Kratzerpotential}
u(\beta)=-{A\over\beta}+{B\over\beta^2},  \qquad A>0
\end{equation}
in $\beta$, while in $\gamma$, we use a harmonic oscillator potential,
$u(\gamma) = c \gamma^2/2$, as in Ref. \cite{Fortunato1}.
Solving the ``radial'' ($\beta$) equation by AIM, we obtain the
energy eigenvalues
\begin{equation}\label{KratzerX5eigenvalue1}
\varepsilon_{n,L}={A^2/4\over\left(n+{1\over2}+\sqrt{{9\over4}+\lambda+B+{L(L+1)\over3}}\right)^2},
\end{equation}
which, in comparison with the result of Ref. \cite{Fortunato1}, contain the additional
term $\lambda$, coming from the procedure of exact separation of variables,
as shown in detail in Appendix A2.
The term $\lambda$ is determined by solving the $\gamma$ equation through AIM, as exhibited in detail in Appendix A2, the final result being
\begin{equation}
\lambda=\left(2c \right)^{1/2} (n_\gamma+1) -K^2/3,
\end{equation}
in agreement with the solution of the $\gamma$ equation given in
 Ref. \cite{Fortunato1}, as demonstrated in Appendix A2.

{\bf Special case:} Taking $B=0$ in Eq. (\ref{Kratzerpotential}),
the Coulomb potential is obtained. The energy eigenvalues are still
given by Eq. (\ref{KratzerX5eigenvalue1}) with $B=0$. Again, the
result differs from the one reported in Ref. \cite{Fortunato1} by
the presence of the $\lambda$ term, due to the exact separation of
variables, as already remarked.

\subsection{Morse potential}

We use the exactly separable form of the potential given in Eq.
(\ref{eq:ES}) again. The Morse potential \cite{Morse} is defined as
\begin{equation}
u(\beta)=e^{-2a(\beta-\beta_e)}-2e^{-a(\beta-\beta_e)}.
\end{equation}
Using the Pekeris approximation \cite{Pekeris} and solving the
$\beta$ equation through AIM (the details are given in Appendix A3),
we obtain the energy eigenvalues
\begin{equation}
\epsilon_{n,L}={\mu
c_0\over\beta_e^2}-\left[{\gamma_1^2\over2\beta_e\gamma_2}-\left(n+{1\over2}\right){\alpha\over\beta_e}\right]^2,
\end{equation}
where
\begin{equation}\label{eq:cs}
c_0=1-{3\over\alpha}+{3\over\alpha^2},\quad
c_1={4\over\alpha}-{6\over\alpha^2},\quad
c_2=-{1\over\alpha}+{3\over\alpha^2}, \quad \alpha = a \beta_e,
\end{equation}
\begin{equation}
\gamma_1^2= 2\beta_e^2-\mu c_1,\quad \gamma_2^2= \beta_e^2+\mu c_2,
\end{equation}
\begin{equation}
\mu={L(L+1)\over3}+2+\lambda.
\end{equation}
$\lambda$ in the last equation comes from the exact separation of variables
and is determined from the $\gamma$ equation. We use the same $\gamma$ potential
(Eq. (\ref{eq:gamma})~) as in the Davidson case (subsection III.A), leading
to the expression for $\lambda=(3C)(n_\gamma+1)-{K^2\over 3}$ as given in Eq. (\ref{eq:lambda}).

\section{$\gamma$-unstable solutions}
\label{DavE5}

In this case, the reduced potential is assumed
to be $\gamma$ independent, $v(\beta,\gamma)=u(\beta)$.
Then the wavefunction is assumed to be of the form \cite{Wilets}
\begin{equation}\label{E5wf}
\psi(\beta,\gamma,\theta_j)=R(\beta)\Phi(\gamma,\theta_j).
\end{equation}
The equation which includes the Euler angles and $\gamma$
 has been solved by B\`es \cite{Bes}.  In this equation, the eigenvalues of
the second-order Casimir operator of SO(5) occur, having the form
$\Lambda=\tau(\tau+3)$, where $\tau$ is the seniority quantum number,
characterizing the irreducible representations of
SO(5) and taking the values $\tau=0,1,2,3,\ldots$ \cite{Rakavy}.

The values of the angular
momentum $L$ are given by the algorithm
\begin{equation}\label{tau}
\tau=3\nu_\Delta+\lambda,\quad\quad \nu_\Delta=0,1,2,\ldots
\end{equation}
\begin{equation}\label{LLambda}
 L=\lambda, \lambda+1, \ldots, 2\lambda-2, 2\lambda
\end{equation}
(with $2\lambda-1$ missing), where $\nu_{\Delta}$ is the missing
quantum number in the reduction $SO(5)\supset SO(3)$ \cite{Rakavy}.
The ground state band levels are determined by $L=2\tau$ and $n=0$.

\subsection{Davidson potential}

The ``radial'' equation, when solved through the AIM, leads to the
energy eigenvalues
\begin{equation}\label{DE5Eigenvalue}
\epsilon_{n,\tau}=2n+1+\left[{9\over 4}+\tau(\tau+3)+\beta_0^4\right]^{1/2},
\end{equation}
the details of the calculation are given in Appendix A4. This result
coincides with those of Ref. \cite{varPLB}.

\emph{\bf Special case:}
Taking $\beta_0=0$ in Eq. (\ref{DE5Eigenvalue}), one gets the simplified expression
\begin{equation}\label{Bohrharmonicoscillator}
\epsilon_{n,\tau}=2n+\tau+{5\over 2},
\end{equation}
which is the 5-dimensional harmonic oscillator solution of Bohr \cite{Bohr}.

\subsection{Kratzer potential}

Solving the ``radial'' equation by AIM (see Appendix A5 for the
details), we obtain the energy eigenvalues
\begin{equation}
\varepsilon_{n,\tau}={A^2/4\over\left(n+{1\over2}+\sqrt{{9\over4}+\tau(\tau+3)+B}\right)^2},
\end{equation}
which coincide with the results of Ref. \cite{Fortunato2}.

{\bf Special case:} For $B=0$, the $\gamma$-unstable solution for the
Coulomb potential is obtained
\begin{equation}
\varepsilon_{n,\tau}={A^2/4\over(n+\tau+2)^2},
\end{equation}
which coincides with the result of Ref. \cite{Fortunato2}.

\subsection{Morse potential}

Using the Pekeris approximation \cite{Pekeris} and AIM (see Appendix
A6 for the details), we obtain the energy eigenvalues
\begin{equation}
\epsilon_{n,\tau}={\nu
c_0\over\beta_e^2}-\left[{\gamma_1^2\over2\beta_e\gamma_2}-\left(n+{1\over2}\right){\alpha\over\beta_e}\right]^2,
\end{equation}
where
\begin{equation}
\gamma_1^2= 2\beta_e^2-\nu c_1,\quad \gamma_2^2= \beta_e^2+\nu c_2,
\end{equation}
\begin{equation}
\nu= \tau(\tau+3)+2,
\end{equation}
with the rest of the quantities given again by Eq. (\ref{eq:cs}).

\section{Numerical results}

In order to test the applicability of the Morse potential in the description of nuclear spectra,
we have fitted all nuclei with mass $A\geq 100$ and $R_{4/2}=E(4)/E(2)<2.6$, for which at least
the $\beta_1$ and $\gamma_1$ bandheads are known, using the $\gamma$-unstable solution of the Morse
potential, which involves two free parameters ($\beta_e$, $a$). Results for 54 nuclei are shown
in Table 1. The quality measure
\begin{equation}\label{eq:e99}
\sigma = \sqrt{ { \sum_{i=1}^n (E_i(exp)-E_i(th))^2 \over
(n-1)E(2_1^+)} },
\end{equation}
used in the rms fits, remains below 1 in most cases.

\begin{figure}
  \includegraphics[width=145mm]{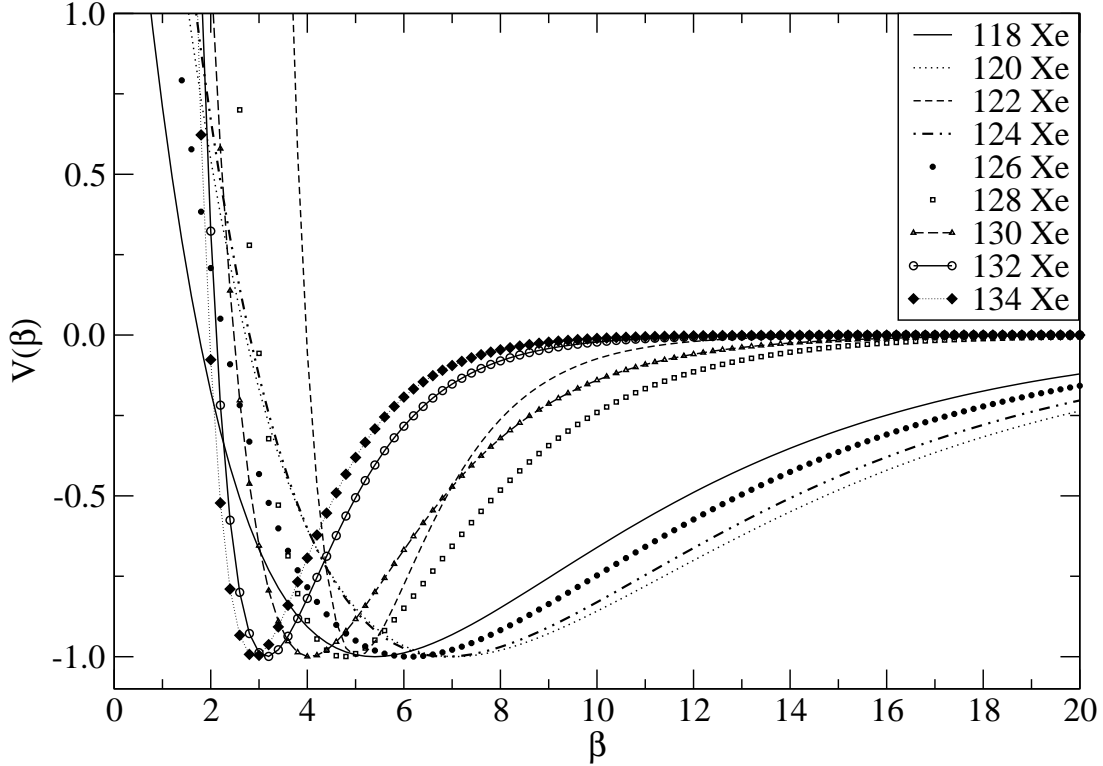}\\
  \caption{Evolution of Morse potential shapes for the $_{54}$Xe isotopes,
  with the parameters given in Table I. See Section V for further discussion.}\label{xe}
\vskip+1cm
\end{figure}

The Morse potentials obtained for the $_{54}$Xe isotopes are shown
in Fig.~1. The evolution of the parameters and the shapes of the
potentials are clear. As one moves from $^{134}$Xe$_{80}$, which is
just below the $N=82$ magic number, to the mid-shell nucleus
$^{120}$Xe$_{66}$, the $\beta_e$ parameter (which is the position of
the minimum of the potential) increases, while the parameter $a$,
which corresponds to the steepness of the potential, decreases. As a
result, one gradually obtains less steep potentials with a minimum
further away from the origin. The trends start to be reversed at
$^{118}$Xe$_{64}$, which is just below mid-shell.

We have also fitted all nuclei
with mass $A\geq 150$ and $R_{4/2}=E(4)/E(2)>2.9$ for which at least
the $\beta_1$ and $\gamma_1$ bandheads are known, using the exactly separable rotational solution of the Morse
potential with $\gamma \approx 0$ (ES-M), which involves three free parameters (the Morse parameters $\beta_e$ and $a$,
as well as the stiffness $C$ of the $\gamma$ potential). All bands are treated on an equal footing, depending on all
three parameters. Results for 45 rare earths and 13 actinides are
shown in Table 2. The quality measure $\sigma$ of Eq. (\ref{eq:e99}), used in the rms fits, remains below 1
in most cases.

\begin{figure}
  \includegraphics[width=145mm]{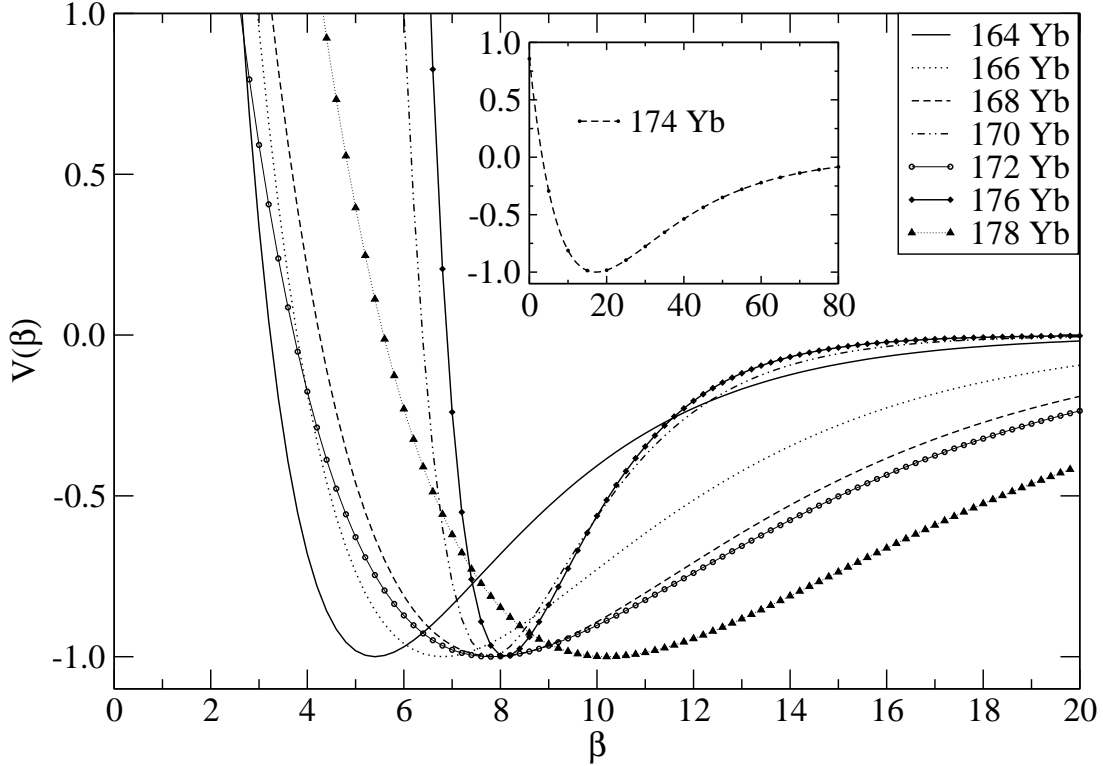}\\
  \caption{Evolution of Morse potential shapes for the $_{70}$Yb isotopes,
  with the parameters given in Table II. See Section V for further discussion. }\label{yb}
  \vskip+1cm
\end{figure}

The Morse potentials obtained for the $_{70}$Yb isotopes are shown
in Fig.~2. The evolution of the parameters and the shapes of the
potentials are again clear. As one moves from $^{164}$Yb$_{94}$ to
the mid-shell nucleus $^{174}$Yb$_{104}$, the $\beta_e$ parameter
(which is the position of the minimum of the $\beta$-potential)
again increases, while the parameter $a$, which corresponds to the
steepness of the $\beta$-potential, again decreases. The $C$
parameter, which is related to the stiffness of the
$\gamma$-potential, increases. As a result, one gradually obtains
less steep $\beta$-potentials with a minimum further away from the
origin, while the $\gamma$-potentials get stiffer at the same time.

A notable exception occurs in the $N=90$ isotones $^{150}$Nd, $^{152}$Sm, $^{154}$Gd, which are known
\cite{CZX5,Kruecken,Tonev}
to be good examples of the X(5) critical point symmetry, along with $^{178}$Os \cite{DewaldOs}. The relative failure
of the Morse potential to describe critical nuclei is expected. The potential at the critical point
is expected to be flat, as the infinite-well potential used in X(5), or to have a little
bump in the middle \cite{JPGreview,Nature}. Microscopic relativistic mean field calculations
\cite{Fossion,Sheng,Vretenar} of potential energy surfaces
support these assumptions. Since the Morse potential cannot imitate a flat potential, with or without
a bump in the middle,  it is expected that it cannot describe these nuclei satisfactorily.

A comparison of the fits of Table 2 to the results provided by the Davidson potential in the exactly separable
$\gamma\approx 0$ case \cite{ESD} (ES-D), which contains two free parameters ($\beta_0$, $c$) instead of three
(see Table 1 of Ref. \cite{ESD}), shows that the extra
parameter extends the region of applicability of the model in most nuclei to higher
angular momenta, largely improving the quality of the fits.

As an example, the spectra of $^{154}$Dy ($\gamma$-unstable case)
and $^{232}$Th (exactly separable rotational case with $\gamma
\approx 0$) are shown in Table 3. The overall agreement between
theory and experiment is very good in both cases. In the theoretical
predictions for the gamma band of $^{154}$Dy, the O(5) degeneracies
are present, limiting the flexibility of the model to agree to
experiment. Spacings within all bands of $^{232}$Th, including the
beta band (in which spacings in X(5) are overpredicted by almost a
factor of two \cite{CZX5,Kruecken,JPGreview}), are reproduced very
accurately.

For the construction of complete level schemes, the calculation of $B(E2)$ transition rates
is required, for which the wave functions are needed. Work in this direction is in progress.

\section{Discussion}

The Bohr Hamiltonian  has been solved with the Morse potential for any angular momentum,
both in the $\gamma$-unstable case and in the exactly separable rotational case with
$\gamma \approx 0$ (in which a harmonic oscillator is used for the $\gamma$ potential),
labelled as ES-M.
The solution has been achieved through the Asymptotic Iteration Method (AIM) and has involved
the Pekeris approximation.
The effectiveness of AIM has been demonstrated by applying it to the $\gamma$-unstable case and
to the exactly separable rotational case with $\gamma\approx 0$ for the Davidson and Kratzer potentials.

Numerical results have been presented for both solutions, including
all relevant medium mass and heavy nuclei for which at least the $\beta_1$ and $\gamma_1$ bandheads are known.
The success of the present solutions in reproducing quite well both the bandeahds of and
the spacings within the ground, $\beta_1$ and $\gamma_1$ bands indicate that a detailed study
of $\gamma_2$ and $\beta_2$ bands within this framework might be fruitful, although the difficulties
in singling out the experimental $\beta_2$ band \cite{Garrett} should be kept in mind.
The influence of the finite depth of the potential is also worth considering in further detail. From
the findings of Ref. \cite{finite}, where the E(5) case was solved for a finite well,
the influence of the finite depth of the potential is expected to show up more clearly in
the higher excited states. Work on the calculation of wave functions and $B(E2)$ transition rates
is in progress.

\section*{Acknoledgments}

I. Boztosun and I. Inci acknowledge the financial support of the
Scientific and Technical Research Council of Turkey
(T\"{U}B\.{I}TAK), under the project number TBAG-107T824 as well as the support
of the Turkish Academy of Sciences (T\"{U}BA-GEB\.{I}P).

\newpage
\centerline{\bf APPENDICES}
\bigskip

\begin{flushleft}
\textbf{A1. Exactly separable $\gamma\approx 0$  solution
for the Davidson potential}
\end{flushleft}
Assuming the reduced potential to be of the form of Eq.
(\ref{eq:ES}), $u(\beta, \gamma) = u(\beta) + {u(\gamma)/\beta^2}$,
and plugging it into the Bohr Hamiltonian of Eq.
(\ref{Bhamiltonian}), we obtain  the ``radial" and $\gamma$
equations \cite{ESD}
\begin{equation} \label{X5Radial}
\left[ -{1\over \beta^4} {\partial \over \partial \beta} \beta^4
{\partial \over \partial \beta} + {L(L+1)\over
3\beta^2}+{\lambda\over\beta^2} +u(\beta) \right] \xi_L(\beta) =
\epsilon \xi_L(\beta),
\end{equation}
\begin{equation}\label{X5Gamma}
\left[-{1\over\sin3\gamma}{\partial\over\partial\gamma}{\sin3\gamma}
{\partial\over\partial\gamma}+{K^2\over4}({1\over\sin^2(\gamma)}
-{4\over3})+u(\gamma)\right]\Gamma_K(\gamma)=\lambda\Gamma_K(\gamma).
\end{equation}
Assuming $u(\gamma)={(3c)^2\gamma^2}$ and expanding
Eq. (\ref{X5Gamma}) in powers of $\gamma$ for $\gamma\simeq0$, we get \cite{ESD}
\begin{equation}\label{X5Gammazero}
\left[-{1\over\gamma}{\partial\over\partial\gamma}
\gamma{\partial\over\partial\gamma}+{K^2\over{4\gamma^2}}+(3c^2){\gamma^2}\right]\Gamma_K(\gamma)=
{\epsilon_\gamma}\Gamma_K(\gamma),
\end{equation}
where $\epsilon_\gamma=\lambda+K^2/3$. The solution of this equation
is given \cite{ESD} in terms of Laguerre polynomials with
\begin{equation}\label{Epsilongamma}
\epsilon_\gamma=(3C)(n_\gamma+1), \qquad C=2c, \quad n_\gamma=0,1,2,3,...
\end{equation}
We now solve the ``radial'' equation by using AIM.
Plugging the Davidson potential of Eq. (\ref{Davidsonpotential}),
$u(\beta)=\beta^2+{\beta_0^4/\beta^2}$,
in Eq. (\ref{X5Radial}), we get
\begin{equation} \label{X5Betahamiltonian}
\left[ -{1\over \beta^4} {\partial \over \partial \beta} \beta^4
{\partial \over \partial \beta} + {{L(L+1)\over
3}+\lambda+\beta_0^4\over\beta^2} +\beta^2 \right] \xi_L(\beta) =
\epsilon \xi_L(\beta).
\end{equation}
Transforming $\xi_L$ into $\chi_L$ by the relation
\begin{equation}\label{X5Betatransformation}
\xi_L(\beta)= \beta^{-2}\chi_L(\beta),
\end{equation}
and plugging it into Eq. (\ref{X5Betahamiltonian}), we obtain
\begin{equation}\label{X5Secondorder}
\chi_L^{''}(\beta)+\left[\epsilon-{{L(L+1)\over3}+\beta_0^4+\lambda+2\over\beta^2}-\beta^2\right]\chi_L(\beta)=0.
\end{equation}
To simplify this equation we define
\begin{equation}\label{eq:muD}
{{L(L+1)\over3}+\lambda+\beta_0^4+2}=\mu{(\mu+1)},
\end{equation}
 obtaining
\begin{equation}\label{SimplyfiedX5secondorder}
\chi_L^{''}(\beta)+\left[\epsilon-{\mu(\mu+1)\over\beta^2}-\beta^2\right]\chi_L(\beta)=0.
\end{equation}
 This second-order differential equation must have a solution of the
form
\begin{equation}\label{Betaoffered}
\chi_L(\beta)=\beta^{\mu+1}e^{-{{\beta^2}\over 2}}f_{n,L}(\beta).
\end{equation}
Using this function in Eq. (\ref{SimplyfiedX5secondorder}),
one can get
\begin{equation}\label{X5Aimform}
f^{''}_{n,L}(\beta)=(2\beta-{{2\mu+2}\over\beta})f^{'}_{n,L}(\beta)
+(2\mu+3-\epsilon)f_{n,L}(\beta).
\end{equation}

From Eq. (\ref{differential}), one can define $\lambda_0(\beta)$ and
$s_0(\beta)$ as follows
\begin{equation}
\lambda_0(\beta)  =  2\beta-{{2\mu+2}\over\beta}, \qquad
s_0(\beta)  =  2\mu+3-\epsilon.
\end{equation}
Then, s$_{k}(x)$ and $\lambda_{k}(x)$ are calculated by the
recurrence relations of Eq. ~(\ref{iteration})
\begin{eqnarray}
\lambda_1(\beta) & = & {4\beta^4-(6\mu+\epsilon+3)\beta^2+4\mu^2+10\mu+6\over\beta^2} \\
s_1(\beta) & = & {2(2\mu+3-\epsilon)(\beta^2-\mu-1)\over\beta} \\
\lambda_2(\beta) & = & {-4\left[(6+13\mu+9\mu^2+2\mu^3)-\beta^2
(3+\epsilon+7\mu+\mu\epsilon+4\mu^2)+\beta^4(4\mu+\epsilon)-2\beta^6
\right]\over\beta^3} \\ s_2(\beta) & = & {(24+52\mu-8\epsilon+36\mu^2-12\epsilon\mu-4\epsilon\mu^2+8\mu^3)\over\beta^2}+ \beta^2(12+8\mu-4\epsilon) \\\nonumber
& - & (3+2\epsilon+20\mu-4\mu\epsilon+12\mu^2-\epsilon^2) \\
 & \vdots & \nonumber
\end{eqnarray}

By applying the termination condition of AIM, Eq. (\ref{quantization}), we get
the energy eigenvalues from the roots of $\Delta_k$ as follows
\begin{equation}
\epsilon_0=\mu+{3\over 2}, \qquad
\epsilon_1 = \mu+{7\over 2}, \qquad
\epsilon_2 = \mu+{11\over 2}, \qquad
\ldots
\end{equation}
while the general expression is
\begin{equation}
\epsilon_{n, \mu}=\mu+{3\over 2}+2n.
\end{equation}
Then, substituting the value of $\mu$ from Eq. (\ref{eq:muD}) yields the energy eigenvalues
\begin{equation}\label{ADX5Eigenvalue}
\epsilon_{n,L}=2n+1+{\left[{9\over4}+{L(L+1)-K^2\over3}+3C(n_\gamma+1)+\beta_0^4\right]}^{1/2},
\end{equation}
which are identical to the ones found in Ref. \cite{ESD}.

\newpage

\begin{flushleft}
\textbf{A2. Exactly separable $\gamma\approx 0$  solution
for the Kratzer potential}
\end{flushleft}

Separation of variables proceeds as in Appendix A1. The ``radial'' equation is
\begin{equation}\label{KratzerX5Radial}
\chi_L^{''}(\beta)+\left[\epsilon-{{L(L+1)\over3}+2+\lambda\over\beta^2}+{A\over\beta}-
{B\over\beta^2}\right]\chi_L(\beta)=0,
\end{equation}
where we have used the wave function of Eq. (\ref{Wavefunction}) and the transformation
$\xi_L(\beta)=\beta^{-2}\chi_L(\beta)$. This equation differs from the corresponding one
of Ref. \cite{Fortunato1} by the term $\lambda$, which comes from the exact separation
of variables and will be determined below from the gamma equation.
Defining a new parameter set
\begin{equation}\label{eq:p}
{L(L+1)\over3}+2+\lambda+B=p(p+1),
\end{equation}
\begin{equation}\label{Kratzertransformatin}
\epsilon=-\varepsilon, \quad 2\beta \sqrt{\varepsilon}=x, \quad
{A\over 2 \sqrt{\varepsilon}}=k,
\end{equation}
Eq. (\ref{KratzerX5Radial}) becomes
\begin{equation}\label{KratzerX5Radiallast}
\chi_L^{''}(x)+\left[-{1\over4}-{p(p+1)\over x^2}+{k\over x}
\right]\chi_L(x)=0.
\end{equation}
This second-order differential equation must have a solution of the form
\begin{equation}
\chi_L(x)=x^{p+1}e^{-x^2\over2}R_{n,L}(x),
\end{equation}
which leads to
\begin{equation}
R_{n,L}^{''}(x)=\left({x-2p-2\over x}\right)R_{n,L}^{'}(x)+\left({p+1-k\over x}\right)R_{n,L}(x).
\end{equation}
According to Eq. (\ref{differential}) of AIM, one then has
\begin{equation}
\lambda_0(x)={x-2p-2\over x}, \qquad s_0(x)={p+1-k\over x}.
\end{equation}
Using the recurrence relations of Eq. (\ref{iteration}),
one can then determine $\lambda_k(x)$ and $s_k(x)$ as
\begin{eqnarray}
\lambda_1(x) & = & {10p+6-3xp-3x-kx+x^2+4p^2\over x^2} \\
s_1(x) & = & {-5p-3+3k+xp-2p^2+x-kx+2kp\over x^2} \\\nonumber
\lambda_2(x) & = & (-24-52p+12x-4x^2-36p^2+6kx+20xp \\
& + & x^3+8p^2x-4px^2+4pkx-8p^3-2kx^2)/x^3 \\\nonumber s_2(x) & = & ((1+p-k)x^2+(3k-4-7p-3p^2+k^2+2pk)x \\ & + & 2(6+13p-6k+9p^2-7kp+2p^3-2kp^2))/ x^3 \\
 & \vdots & \nonumber
\end{eqnarray}

After determining $\Delta_k(x)$ by using the termination condition
Eq. (\ref{quantization}), the energy eigenvalues are found from the roots
of $\Delta_k(x)$ as follows
\begin{equation}
k_0=p+1, \quad k_1=p+2, \quad k_2=p+3,\quad\ldots,
\end{equation}
generalized into
\begin{equation}
k_n=p+1+n.
\end{equation} From
Eqs.  (\ref{eq:p}) and (\ref{Kratzertransformatin}), one then obtains
\begin{equation}\label{KratzerX5eigenvalue}
\varepsilon_{n,L}={A^2/4\over\left(n+{1\over2}+\sqrt{{9\over4}+\lambda+B+{L(L+1)\over3}}\right)^2}.
\end{equation}

To find $\lambda$, we have to solve the $\gamma$ equation of Eq.
(\ref{X5Gamma}), using \cite{Fortunato1} $u(\gamma)={c\gamma^2/2}$.
Expanding around $\gamma=0$ and taking
$\epsilon_{\gamma}=\lambda+{K^2/3}$, one obtains
\begin{equation}\label{Kratzergamma}
\left[-{1\over\gamma}{\partial\over\partial\gamma}{\gamma}
{\partial\over\partial\gamma}+{K^2\over4\gamma^2}+{c\gamma^2\over2}\right]\Gamma_K(\gamma)=\epsilon_{\gamma}\Gamma_K(\gamma),
\end{equation}
which, through the transformation  $\Gamma_K(\gamma)=\gamma^{-1/2}\xi_K(\gamma)$, leads to
\begin{equation}
\xi_K^{''}(\gamma)+\left[\epsilon_{\gamma}+{(1-K^2)/4\over\gamma^2}-{c\gamma^2\over2}\right]\xi_K(\gamma)=0.
\end{equation}
Defining
\begin{equation} \label{eq:mu}
{\left({c\over2}\right)^{1/4}\gamma=y}, \qquad \mu(\mu+1)=(K^2-1)/4, \qquad
\varepsilon_{\gamma}=({c\over2})^{-1/2}\epsilon_{\gamma},
\end{equation}
this equation is brought into the form
\begin{equation}\label{Kratzergammay}
\xi_K^{''}(y)+\left[\varepsilon_{\gamma}-{\mu(\mu+1)\over
y^2}-y^2\right]\xi_K(y)=0,
\end{equation}
which is suitable for solving through AIM,
by considering a solution of the form
\begin{equation}\label{Kratzergammaysol}
\xi_K(y)=y^{\mu+1}e^{-{y^2\over2}}G_{m,K}(y).
\end{equation}
Following the same procedure as above, we obtain the differential equation in
the form
\begin{equation}
G^{''}_{m,K}(y)=\left({2y^2-2\mu-2\over y}\right)G^{'}_{m,K}(y)+
\left({2y\mu+3y-y\varepsilon_{\gamma}\over y}\right)G_{m,K}(y).
\end{equation}
Comparison with Eq. (\ref{differential}) leads to
\begin{equation}
\lambda_0(y)={2y^2-2\mu-2\over y}, \qquad
s_0(y)={2\mu+3-\varepsilon_{\gamma}}.
\end{equation}
By using the recurrence relations of Eq. (\ref{iteration}) and
the termination condition of Eq. (\ref{quantization}), one can find
the energy eigenvalues to be
\begin{equation}
(\varepsilon_{\gamma})_0=2\mu+3, \quad (\varepsilon_{\gamma})_1=2\mu+7, \quad
(\varepsilon_{\gamma})_2=2\mu+11, \quad \ldots,
\end{equation}
and the generalized form is
\begin{equation}
(\varepsilon_{\gamma})_{\mu,m}=2\mu+3+4m.
\end{equation}
Using $\mu$ from Eq. (\ref{eq:mu}), this becomes
\begin{equation}
(\epsilon_{\gamma})_{K,m}=\left(c\over2\right)^{1/2}\left[4m+2+K\right],
\end{equation}
leading to
\begin{equation}
\lambda=\left(c\over2\right)^{1/2}\left[4m+2+K\right]-K^2/3.
\end{equation}
This result agrees with Ref. \cite{Fortunato1}, with $K= 2n_\gamma -4m$,
$m=0$, 1, \dots, $n_\gamma$, as stated there. It is also in full agreement with
the results of Ref. \cite{ESD}, taking into account the different coefficients
of $\gamma^2$ in $u(\gamma)$.

\begin{flushleft}
\textbf{A3. Exactly separable $\gamma\approx 0$  solution
for the Morse potential}
\end{flushleft}

Separation of variables again proceeds as in Appendix A1. Using the transformation
$\xi_L(\beta)=\beta^{-2}\chi_L(\beta)$, the ``radial'' equation becomes
\begin{equation}\label{Morse1}
\chi^{''}_L(\beta)+\left[\epsilon-{{L(L+1)\over3}+\lambda+2\over\beta^2}-
e^{-2a(\beta-\beta_e)}+2e^{-a(\beta-\beta_e)}\right]\chi_L(\beta)=0.
\end{equation}
Defining
\begin{equation}
x={\beta-\beta_e\over\beta_e}, \quad \alpha={a\beta_e}, \quad
\beta_e^2\epsilon=\varepsilon, \quad{L(L+1)\over3}+2+\lambda=\mu,
\end{equation}
the ``radial'' equation becomes
\begin{equation}\label{Morse2}
\chi^{''}_L(x)+\left[\varepsilon-{\mu\over(1+x)^2}-
\beta_e^2e^{-2\alpha x}+2\beta_e^2e^{-\alpha x}\right]\chi_L(x)=0.
\end{equation}
We now apply the Pekeris approximation \cite{Pekeris}. Renaming
${\mu\over(1+x)^2}$ as $u_L(x)$ and expanding in a series
around $x=0$ we get
\begin{equation}\label{uL}
u_L(x)=\mu(1-2x+3x^2-4x^3+\ldots).
\end{equation}
In the exponential form, $u_L(x)$ can be written as
\begin{equation}
\tilde{u}_L(x)=\mu(c_0+c_1e^{-\alpha x}+c_2e^{-2\alpha x}+\ldots).
\end{equation}
Expanding also this equation in a series around $x=0$, we get
\begin{equation}\label{uLtilde}
\tilde{u}_L(x)=\mu\left(c_0+c_1+c_2-[c_1+2c_2]\alpha
x+[{c_1\over2}+2c_2]\alpha^2x^2+\ldots\right).
\end{equation}
Comparing Eqs. (\ref{uL}) with Eq. (\ref{uLtilde}), one can now determine the $c_i$ coefficients
\begin{equation}
c_0=1-{3\over\alpha}+{3\over\alpha^2},\quad
c_1={4\over\alpha}-{6\over\alpha^2},\quad
c_2=-{1\over\alpha}+{3\over\alpha^2}.
\end{equation}
Returning to the ``radial''  equation
\begin{equation}\label{Morseaim}
\chi^{''}_L(x)+\left[\varepsilon-\mu\left(c_0+c_1e^{-\alpha
x}+c_2e^{-2\alpha x}\right)- \beta_e^2e^{-2\alpha
x}+2\beta_e^2e^{-\alpha x}\right]\chi_L(x)=0,
\end{equation}
and by using the ansatz
\begin{equation}\label{ansatz}
\varepsilon-\mu c_0=-\rho^2,\quad 2\beta_e^2-\mu
c_1=\gamma_1^2,\quad \beta_e^2+\mu c_2=\gamma_2^2,
\end{equation}
we get
\begin{equation}\label{Morseaim1}
\chi^{''}_L(x)+\left[-\rho^2+\gamma_1^2e^{-\alpha
x}-\gamma_2^2e^{-2\alpha x}\right]\chi_L(x)=0.
\end{equation}
Rewriting this equation by using the new variable $y=e^{-\alpha x}$, we obtain
\begin{equation}\label{Morseaim2}
\chi^{''}_L(y)+{1\over
y}\chi^{'}_L(y)+\left[-{\rho^2\over\alpha^2y^2}+{\gamma_1^2\over\alpha^2y}-{\gamma_2^2\over\alpha^2}\right]\chi_L(y)=0.
\end{equation}
Inserting a wave function of the form
\begin{equation}
\chi_L(y)=y^{\rho\over\alpha}e^{-{\gamma_2\over\alpha}y}R_{n,L}(y),
\end{equation}
the second order differential equation becomes
\begin{equation}
R_{n,L}^{''}(y)=\left({2\gamma_2\alpha y-2\alpha\rho-\alpha^2\over
\alpha^2y}\right)R_{n,L}^{'}(y)+\left(2\rho\gamma_2+\alpha\gamma_2-\gamma_1^2\over\alpha^2y\right)R_{n,L}(y).
\end{equation}
Comparison with Eq. (\ref{newdifferential}) leads to the identifications
\begin{eqnarray}
\tau(y) & = & 2\gamma_2 \alpha y-2\rho-\alpha, \nonumber \\
\sigma(y) & = & \alpha^2 y, \nonumber \\
\Omega_n & = & 2\gamma_2\rho+\alpha\gamma_2-\gamma_1^2.
\end{eqnarray} From
Eq.(\ref{gamman}), we get
\begin{equation}
2\gamma_2\rho+\alpha\gamma_2-\gamma_1^2=-n(2\gamma_2\alpha),
\end{equation}
while the generalized form is written as
\begin{equation}
\rho_{n,L}={\gamma_1^2\over2\gamma_2}-{(n+{1\over2})\alpha}.
\end{equation}
Using Eq. (\ref{ansatz}), one then obtains
\begin{equation}
\epsilon_{n,L}={\mu
c_0\over\beta_e^2}-\left[{\gamma_1^2\over2\beta_e\gamma_2}-\left(n+{1\over2}\right){\alpha\over\beta_e}\right]^2.
\end{equation}

\begin{flushleft}
\textbf{A4. $\gamma$-unstable solution for the Davidson potential}
\end{flushleft}

In this case, the reduced potential depends only on $\beta$.
Using the wave function of Eq. (\ref{E5wf}),
the relevant ``radial'' equation becomes
\begin{equation}\label{E5Wavefunction}
\chi_{\tau}^{''}(\beta)+\left[\epsilon-\beta^2-{p(p+1))\over\beta^2}\right]\chi_{\tau}(\beta)=0,
\end{equation}
where $\chi_{\tau}(\beta)=\beta^{-2} R_{\tau}(\beta)$
 and
 \begin{equation}\label{eq:pp}
 p(p+1)=\tau(\tau+3)+\beta_0^4+2.
 \end{equation}
 For this
differential equation, one looks for a solution of the form
\begin{equation}
\chi_{\tau}(\beta)=\beta^pe^{-\beta^2\over2}f_{n,\tau}(\beta).
\end{equation}
Using this function in Eq. (\ref{E5Wavefunction}), we get a
differential equation, which is similar to Eq. (\ref{differential})
\begin{equation}
f_{n,\tau}^{''}(\beta)=\left({2\beta^3-2p\beta\over\beta^2}\right)f_{n,\tau}^{'}(\beta)+\left({\beta^2+2p+2p\beta^2
-\epsilon\beta^2\over\beta^2}\right)f_{n,\tau}(\beta).
\end{equation}
Comparing to Eq. (\ref{differential}),  we identify
\begin{equation}
\lambda_0(\beta)={2\beta^3-2p\beta\over\beta^2}, \qquad
s_0(\beta)={\beta^2+2p+2p\beta^2 -\epsilon\beta^2\over\beta^2}.
\end{equation}
Then the recurrence relations of Eq. (\ref{iteration}) give
$\lambda_k(\beta)$ and $s_k(\beta)$, while by using the termination
relations of Eq. (\ref{quantization}), we can obtain the energy
eigenvalues
\begin{equation}
\epsilon_0=p+{3\over 2}, \quad \epsilon_1=p+{7\over 2}, \quad \epsilon_2=p+{11\over 2}, \quad\ldots,
\end{equation}
which give the generalized form
\begin{equation}
\epsilon_{n,p}=2n+p+{3\over 2}.
\end{equation}
Substituting in this expression the value of $p$ from Eq. (\ref{eq:pp}), one gets
\begin{equation}\label{ADE5Eigenvalue}
\epsilon_{n,\tau}=2n+1+\left[{9\over 4}+\tau(\tau+3)+\beta_0^4\right]^{1/2},
\end{equation}
in agreement with the result obtained in Ref. \cite{varPLB}.

\begin{flushleft}
\textbf{A5. $\gamma$-unstable solution for the Kratzer potential}
\end{flushleft}

Using wavefunctions of the form
$\Psi(\beta,\gamma,\theta_i)=\chi(\beta) \Phi(\gamma,\theta_i)$,
the ``radial'' equation becomes
\begin{equation}
\left[-{1\over\beta^4}{\partial\over\partial\beta}\beta^4{\partial\over\partial\beta}
+{\tau(\tau+3)\over\beta^2}-{A\over\beta}+{B\over\beta^2}\right]\chi(\beta)=\epsilon\chi(\beta),
\end{equation}
in agreement with Ref. \cite{Fortunato2}. Substituting
$\chi(\beta)=\beta^{-2}\xi(\beta)$, this becomes
\begin{equation}
\xi^{''}(\beta)+\left[\epsilon-{\tau(\tau+3)+B+2\over\beta^2}+{A\over\beta}\right]\xi(\beta)=0.
\end{equation}
Using the parameter set
\begin{equation}\label{tautau}
\epsilon=-\varepsilon, \quad 2\beta \sqrt{\varepsilon}=y, \quad
{A\over 2 \sqrt{\varepsilon}}=k, \quad {\tau(\tau+3)+B+2}=\nu(\nu+1),
\end{equation}
the differential equation becomes
\begin{equation}
\xi_{\tau}^{''}(y)+\left[-{1\over4}+{k\over y}-{\nu(\nu+1)\over y^2}
\right]\xi_{\tau}(y)=0.
\end{equation}
Assuming that this equation has a solution of the form
\begin{equation}
\xi_{\tau}(y)=y^{\nu+1}e^{-{y\over2}}R_{n,\tau}(y),
\end{equation}
we bring it into the form
\begin{equation}
R^{''}_{n,\tau}(y)=\left({y-2\nu-2\over
y}\right)R^{'}_{n,\tau}(y)+\left({\nu+1-k\over
y}\right)R_{n,\tau}(y).
\end{equation}
Comparison with Eq. (\ref{differential}) then provides
\begin{equation}
\lambda_0(y)={y-2\nu-2\over y}, \qquad
s_0(y)={\nu+1-k\over y}.
\end{equation}
By using the recurrence relations of Eq. (\ref{iteration}) and the termination conditions
of Eq. (\ref{quantization}), one then obtains the energy eigenvalues from the roots of $\Delta_i(y)$ as given
below
\begin{equation}
k_0=\nu+1, \quad k_1=\nu+2, \quad k_2=\nu+3, \quad\ldots.
\end{equation}
These are generalized into
\begin{equation}
k_n=\nu+1+n.
\end{equation} From
Eq. (\ref{tautau}), one then obtains the energy eigenvalues
\begin{equation}
\varepsilon_{n,\tau}={A^2/4\over\left(n+{1\over2}+\sqrt{{9\over4}+\tau(\tau+3)+B}\right)^2},
\end{equation}
in agreement with Ref. \cite{Fortunato2}.

\begin{flushleft}
\textbf{A6. $\gamma$-unstable solution for the Morse potential}
\end{flushleft}

Using a wave function of the form of Eq. (\ref{E5wf}), the ``radial'' equation is
\begin{equation}
\left[-{1\over\beta^4}{\partial\over\partial\beta}\beta^4{\partial\over\partial\beta}
+{\tau(\tau+3)\over\beta^2}+u(\beta)\right]\xi(\beta)=\epsilon\xi(\beta)
\end{equation}
Taking $\xi(\beta)=\beta^{-2}\chi(\beta)$ and
\begin{equation}\label{eq:taunu}
\tau(\tau+3)+2=\nu,
\end{equation}
this equation becomes
\begin{equation}\label{E5Morseham1}
\chi^{''}(\beta)+\left[\epsilon-{\nu\over\beta^2}-u(\beta)\right]\chi(\beta)=0.
\end{equation}
Using the parametrization
\begin{equation}
x={\beta-\beta_e\over\beta_e},\quad\alpha=a\beta_e,\quad
\varepsilon=\beta_e^2\epsilon,
\end{equation}
one obtains
\begin{equation}\label{E5Morseham}
\chi^{''}(x)+\left[\varepsilon-{\nu\over(1+x)^2}-\beta_e^2e^{-2\alpha
x}+2\beta_e^2e^{-\alpha x}\right]\chi(x)=0.
\end{equation}
Applying now the Pekeris approximation \cite{Pekeris} as in Appendix A3,
we replace $1/(1+x)^2$ by its approximate expression, obtaining
\begin{equation}
\chi^{''}(x)+\left[\varepsilon-{\nu(c_0+c_1e^{-\alpha
x}+c_2e^{-2\alpha x})}-\beta_e^2e^{-\alpha x}+2\beta_e^2e^{-\alpha
x}\right]\chi(x)=0.
\end{equation}
Using the parametrization
\begin{equation}\label{eps}
\varepsilon-\nu c_0=-K^2,\quad 2\beta_e^2-\nu c_1=\gamma_1^2,\quad
\beta_e^2+\nu c_2=\gamma_2^2,
\end{equation}
the differential equation is brought into the form
\begin{equation}\label{MorseE5aim1}
\chi^{''}(x)+\left[-K^2+\gamma_1^2e^{-\alpha
x}-\gamma_2^2e^{-2\alpha x}\right]\chi(x)=0.
\end{equation}
Introducing a new variable $y=e^{-\alpha x}$, one has
\begin{equation}\label{MorseE5aim2}
\chi^{''}_{\tau}(y)+{1\over
y}\chi^{'}_{\tau}(y)+\left[-{K^2\over\alpha^2y^2}+{\gamma_1^2\over\alpha^2y}-{\gamma_2^2\over\alpha^2}\right]\chi_{\tau}(y)=0.
\end{equation}
Inserting a wave function of the form
\begin{equation}
\chi_{\tau}(y)=y^{K\over\alpha}e^{-{\gamma_2\over\alpha}y}f_{n,\tau}(y),
\end{equation}
the differential equation becomes
\begin{equation}
f_{n,\tau}^{''}(y)=\left({2\gamma_2\alpha y-2\alpha K-\alpha^2\over
\alpha^2y}\right)f_{n,\tau}^{'}(y)+\left(2K\gamma_2+\alpha\gamma_2-\gamma_1^2\over\alpha^2y\right)f_{n,\tau}(y).
\end{equation}

Comparison with Eq. (\ref{newdifferential}) leads to
\begin{equation}
\tau(y)=2\gamma_2y-2K_n-\alpha, \qquad
\sigma(y)=\alpha^2 y, \qquad
\Omega_n=2\gamma_2K_n+\alpha\gamma_2-\gamma_1^2.
\end{equation}
Using Eq.(\ref{gamman}), one then has
\begin{equation}
K_{n,\tau}={\gamma_1^2\over2\gamma_2}-{\left(n+{1\over2}\right)\alpha}.
\end{equation} From
Eq. (\ref{eps}), we obtain the energy eigenvalues
\begin{equation}
\epsilon_{n,\tau}={\nu
c_0\over\beta_e^2}-\left[{\gamma_1^2\over2\beta_e\gamma_2}-\left(n+{1\over2}\right){\alpha\over\beta_e}\right]^2.
\end{equation}

\begin{table}

\caption{Comparison of theoretical predictions of the
$\gamma$-unstable Bohr Hamiltonian with Morse potential to
experimental data \cite{NDS} of nuclei with $A\geq 100$, $R_{4/2}
\leq 2.6$,  and known $0_2^+$ and $2_{\gamma}^+$ states.  The
$R_{4/2}=E(4_1^+)/E(2_1^+)$ ratios, as well as the $\beta$ and
$\gamma$ bandheads, normalized to the $2_1^+$ state and labelled by
$R_{0/2}=E(0_{\beta}^+)/E(2_1^+)$ and
$R_{2/2}=E(2_{\gamma}^+)/E(2_1^+)$ respectively, are shown. The
angular momenta of the highest levels of the ground state, $\beta$
and $\gamma$ bands included in the rms fit are labelled by $L_g$,
$L_\beta$, and $L_\gamma$ respectively, while $n$ indicates the
total number of levels involved in the fit and $\sigma$ is the
quality measure of Eq. (\ref{eq:e99}). See Section V for further
discussion. }

\bigskip

\begin{tabular}{ r r r r  r r r r  r r r r r r}
\hline nucleus & $R_{4/2}$ & $R_{4/2}$ & $R_{0/2}$& $R_{0/2}$
&$R_{2/2}$ & $R_{2/2}$ & $\beta_0$ & $a$ &
$L_g$ & $L_\beta$ & $L_\gamma$ & $n$ & $\sigma$ \\
        & exp &  th  & exp & th  & exp & th &  &  &  &  &  &   &  \\

\hline

$^{98}$Ru  & 2.14 & 2.25 & 2.0 &  2.4 &  2.2 &  2.3 & 3.84 & 0.44
& 18 &   0 &   4 & 12 & 0.659 \\
$^{100}$Ru & 2.27 & 2.29 & 2.1 &  2.8 &  2.5 &  2.3 & 4.43 & 0.36
& 28 &   0 &   4 & 17 & 0.291 \\
$^{102}$Ru & 2.33 & 2.25 & 2.0 &  2.3 &  2.3 &  2.2 & 3.78 & 0.42
& 16 &   0 &   5 & 12 & 0.341 \\
$^{104}$Ru & 2.48 & 2.33 & 2.8 &  2.9 &  2.5 &  2.3 & 7.57 & 0.10
&  8 &  2  &   8 & 12 & 0.433 \\

$^{102}$Pd & 2.29 & 2.28 & 2.9 & 2.6 & 2.8 & 2.3 & 4.30 & 0.34
& 26 &  4  &  4 & 18 & 0.219 \\
$^{104}$Pd & 2.38 & 2.28 & 2.4 & 2.6 & 2.4 & 2.3 & 4.15 & 0.41
& 18 &  2  &  4 & 13 & 0.300 \\
$^{106}$Pd & 2.40 & 2.26 & 2.2 & 2.4 & 2.2 & 2.3 & 3.93 & 0.43
& 16 &  4  &  5 & 14 & 0.343 \\
$^{108}$Pd & 2.42 & 2.28 & 2.4 & 2.5 & 2.1 & 2.3 & 4.36 & 0.30
& 14 &  4  &  4 & 12 & 0.313 \\
$^{110}$Pd & 2.46 & 2.29 & 2.5 & 2.0 & 2.2 & 2.3 & 4.01 & 0.26
& 12 &  10 &  4 & 14 & 0.338 \\
$^{112}$Pd & 2.53 & 2.33 & 2.6 & 2.5 & 2.1 & 2.3 & 4.11 & 0.60
&  6 &   0 &  3 &  5 & 0.485 \\
$^{114}$Pd & 2.56 & 2.31 & 2.6 & 2.9 & 2.1 & 2.3 & 5.12 & 0.24
& 16 &   0 & 11 & 18 & 0.727 \\
$^{116}$Pd & 2.58 & 2.34 & 3.3 & 3.5 & 2.2 & 2.3 & 7.44 & 0.13
& 16 &   0 &  9 & 16 & 0.626 \\

$^{106}$Cd & 2.36 & 2.37 & 2.8 & 2.9 & 2.7 & 2.4 & 4.45 & 0.62
& 12 & 0 & 2 &  7 & 0.196 \\
$^{108}$Cd & 2.38 & 2.26 & 2.7 & 2.5 & 2.5 & 2.3 & 3.97 & 0.43
& 18 & 0 & 5 & 13 & 0.688 \\
$^{110}$Cd & 2.35 & 2.24 & 2.2 & 2.2 & 2.2 & 2.2 & 3.66 & 0.47
& 16 & 6 & 5 & 15 & 0.269 \\
$^{112}$Cd & 2.29 & 2.23 & 2.0 & 2.1 & 2.1 & 2.2 & 3.55 & 0.50
& 12 & 8 &11 & 20 & 0.542 \\
$^{114}$Cd & 2.30 & 2.21 & 2.0 & 2.0 & 2.2 & 2.2 & 3.43 & 0.51
& 14 & 4 & 3 & 11 & 0.359 \\
$^{116}$Cd & 2.38 & 2.29 & 2.5 & 2.7 & 2.4 & 2.3 & 4.10 & 0.47
& 14 & 2 & 3 & 10 & 0.408 \\
$^{118}$Cd & 2.39 & 2.29 & 2.6 & 2.7 & 2.6 & 2.3 & 4.11 & 0.47
& 14 & 0 & 3 &  9 & 0.313 \\
$^{120}$Cd & 2.38 & 2.28 & 2.7 & 2.6 & 2.6 & 2.3 & 4.09 & 0.44
& 16 & 0 & 2 &  9 & 0.379 \\

\hline
\end{tabular}
\end{table}

\begin{table}
\setcounter{table}{0} \caption{(continued)}

\bigskip

\begin{tabular}{ r r r r  r r r r  r r r r r r}
\hline nucleus & $R_{4/2}$ & $R_{4/2}$ & $R_{0/2}$& $R_{0/2}$
&$R_{2/2}$ & $R_{2/2}$ & $\beta_0$ & $a$ &
$L_g$ & $L_\beta$ & $L_\gamma$ & $n$ & $\sigma$ \\
        & exp &  th  & exp & th  & exp & th &  &  &  &  &    & &  \\

\hline

$^{118}$Xe & 2.40 & 2.31 & 2.5 & 2.7 & 2.8 & 2.3 & 5.40 & 0.19
& 16 & 4 &10 & 19 & 0.343 \\
$^{120}$Xe & 2.47 & 2.35 & 2.8 & 3.8 & 2.7 & 2.4 & 7.05 & 0.16
& 26 & 4 & 9 & 23 & 0.652 \\
$^{122}$Xe & 2.50 & 2.43 & 3.5 & 3.5 & 2.5 & 2.4 & 5.03 & 0.66
& 10 & 0 & 9 & 13 & 0.501 \\
$^{124}$Xe & 2.48 & 2.35 & 3.6 & 3.8 & 2.4 & 2.4 & 6.88 & 0.17
& 20 & 2 &11 & 21 & 0.562 \\
$^{126}$Xe & 2.42 & 2.33 & 3.4 & 3.2 & 2.3 & 2.3 & 6.12 & 0.18
& 12 & 4 & 9 & 16 & 0.576 \\
$^{128}$Xe & 2.33 & 2.33 & 3.6 & 3.3 & 2.2 & 2.3 & 4.74 & 0.39
& 10 & 2 & 7 & 12 & 0.522 \\
$^{130}$Xe & 2.25 & 2.27 & 3.3 & 2.6 & 2.1 & 2.3 & 4.05 & 0.44
& 14 & 0 & 5 & 11 & 0.476 \\
$^{132}$Xe & 2.16 & 2.17 & 2.8 & 1.4 & 1.9 & 2.2 & 3.16 & 0.66
&  6 & 0 & 5 &  7 & 0.731 \\
$^{134}$Xe & 2.04 & 2.11 & 1.9 & 1.0 & 1.9 & 2.1 & 2.91 & 0.74
&  6 & 0 & 5 &  7 & 0.753 \\

$^{130}$Ba & 2.52 & 2.46 & 3.3 & 3.3 & 2.5 & 2.5 & 5.58 & 0.77
& 12 & 0 & 6 & 11 & 0.416 \\
$^{132}$Ba & 2.43 & 2.29 & 3.2 & 2.7 & 2.2 & 2.3 & 4.63 & 0.29
& 14 & 0 & 8 & 14 & 0.609 \\
$^{134}$Ba & 2.32 & 2.26 & 2.9 & 2.4 & 1.9 & 2.3 & 3.82 & 0.50
&  8 & 0 & 4 &  7 & 0.483 \\
$^{136}$Ba & 2.28 & 2.18 & 1.9 & 1.7 & 1.9 & 2.2 & 3.24 & 0.60
&  6 & 0 & 2 &  4 & 0.454 \\
$^{142}$Ba & 2.32 & 2.44 & 4.3 & 4.3 & 4.0 & 2.4 & 5.45 & 0.60
& 14 & 0 & 2 &  8 & 0.605 \\

$^{134}$Ce & 2.56 & 2.37 &  3.7 &  4.1 &  2.4 & 2.4 &  6.00 & 0.27 &
34 & 2 & 8 & 25 & 0.502 \\
$^{136}$Ce & 2.38 & 2.24 &  1.9 &  2.2 &  2.0 & 2.2 &  3.66 & 0.47 &
16 & 0 & 3 & 10 & 0.618 \\
$^{138}$Ce & 2.32 & 2.23 &  1.9 &  2.1 &  1.9 & 2.2 &  3.55 & 0.50 &
14 & 0 & 2 &  8 & 1.308 \\

$^{140}$Nd & 2.33 & 2.19 &  1.8 &  1.7 & 1.9 & 2.2 & 3.27 & 0.60 &
 6 & 0 & 2 &  4 & 0.265 \\
$^{148}$Nd & 2.49 & 2.32 &  3.0 &  2.9 & 4.1 & 2.3 & 6.40 & 0.14 &
12 & 8 & 4 & 13 & 0.810 \\

$^{140}$Sm & 2.35 & 2.38 & 1.9 & 1.9 & 2.7 & 2.4 & 4.20 & 0.77 &
 8 & 0 & 2 &  5 & 0.153 \\
$^{142}$Sm & 2.33 & 2.20 & 1.9 & 1.7 & 2.2 & 2.2 & 3.33 & 0.61 &
 8 & 0 & 2 &  5 & 0.173 \\

$^{142}$Gd & 2.35 & 2.28 & 2.7 & 2.7 & 1.9 & 2.3 & 4.17 & 0.42 &
16 & 0 & 2 &  9 & 0.188 \\
$^{144}$Gd & 2.35 & 2.36 & 2.5 & 2.5 & 2.5 & 2.4 & 4.24 & 0.65 &
 6 & 0 & 2 &  4 & 0.102 \\
$^{152}$Gd & 2.19 & 2.26 & 1.8 & 2.4 & 3.2 & 2.3 & 3.93 & 0.40 &
16 & 10& 7 & 19 & 0.436 \\

$^{154}$Dy  & 2.23 & 2.28 & 2.0 & 2.7 & 3.1 & 2.3 & 4.22 & 0.38 &
26 & 10 & 7 & 24 & 0.371 \\

$^{156}$Er  & 2.32 & 2.31 & 2.7 & 3.1 & 2.7 & 2.3 & 4.75 & 0.34 &
20 & 4 & 5 & 16 & 0.374 \\

\hline
\end{tabular}
\end{table}

\begin{table}
\setcounter{table}{0} \caption{(continued)}

\bigskip

\begin{tabular}{ r r r r  r r r r  r r r r r r}
\hline nucleus & $R_{4/2}$ & $R_{4/2}$ & $R_{0/2}$& $R_{0/2}$
&$R_{2/2}$ & $R_{2/2}$ & $\beta_0$ & $a$ &
$L_g$ & $L_\beta$ & $L_\gamma$ & $n$ & $\sigma$ \\
        & exp &  th  & exp & th  & exp & th &  &  &  &  &    & &  \\

\hline

$^{186}$Pt & 2.56 & 2.34 & 2.5 & 1.7 & 3.2 & 2.3 & 6.18 & 0.07 &
26 & 6 & 10 & 25 & 1.070 \\
$^{188}$Pt & 2.53 & 2.45 & 3.0 & 3.3 & 2.3 & 2.5 & 5.45 & 0.75 &
16 & 2 & 4 & 12 & 0.356 \\
$^{190}$Pt & 2.49 & 2.34 & 3.1 & 3.6 & 2.0 & 2.3 & 5.08 & 0.35 &
18 & 2 & 6 & 15 & 0.566 \\
$^{192}$Pt & 2.48 & 2.35 & 3.8 & 3.7 & 1.9 & 2.3 & 6.42 & 0.19 &
10 & 0 & 8 & 12 & 0.681 \\
$^{194}$Pt & 2.47 & 2.35 & 3.9 & 3.6 & 1.9 & 2.3 & 7.28 & 0.14 &
10& 4 & 5 & 11 & 0.657 \\
$^{196}$Pt & 2.47 & 2.32 & 3.2 & 2.9 & 1.9 & 2.3 & 6.26 & 0.15 &
10 & 2 & 6 & 11 & 0.627 \\
$^{198}$Pt & 2.42 & 2.25 & 2.2 & 2.3 & 1.9 & 2.3 & 3.87 & 0.39 &
 6 & 2 & 4 & 7  & 0.374 \\
$^{200}$Pt & 2.35 & 2.20 & 2.4 & 1.7 & 1.8 & 2.2 & 3.31 & 0.59 &
 4 & 0 & 4 &  5 & 0.676 \\

\hline
\end{tabular}
\end{table}

\begin{table}

\caption{Comparison of theoretical predictions of the exactly
separable Morse model [ES-M] to experimental data \cite{NDS} of
rare earth and actinides with $A \geq 150$,  $R_{4/2} > 2.9$, and known $0_2^+$
and $2_{\gamma}^+$ states. Data for $^{228}$Ra come from Ref. \cite{Cocks}.
 The $R_{4/2}=E(4_1^+)/E(2_1^+)$
ratios, as well as the $\beta$ and $\gamma$ bandheads, normalized to
the $2_1^+$ state and labelled by $R_{0/2}=E(0_{\beta}^+)/E(2_1^+)$
and $R_{2/2}=E(2_{\gamma}^+)/E(2_1^+)$ respectively, are shown. The
angular momenta of the highest levels of the ground state, $\beta$
and $\gamma$ bands included in the rms fit are labelled by $L_g$,
$L_\beta$, and $L_\gamma$ respectively, while $n$ indicates the
total number of levels involved in the fit and $\sigma$ is the
quality measure of Eq. (\ref{eq:e99}). See Section V  for further
discussion. }

\bigskip

\begin{tabular}{ l r r r  r r r r r r r r r r r}
\hline nucleus & $R_{4/2}$ & $R_{4/2}$ & $R_{0/2}$& $R_{0/2}$
&$R_{2/2}$ & $R_{2/2}$ & $\beta_0$ & $C$   & $a$ &
$L_g$ & $L_\beta$ & $L_\gamma$ & $n$ & $\sigma$ \\
        & exp &  th  & exp & th  & exp & th &  &  &  & & &  &   &  \\

\hline

$^{150}$Nd & 2.93 & 3.21 & 5.2 & 6.2 & 8.2 & 8.2 & 5.2 & 6.0 &  0.41
& 14 &   6 &   4 & 13 & 1.129 \\

$^{152}$Sm & 3.01 & 3.22 & 5.6 & 6.9 & 8.9 & 9.5 & 5.3 & 7.0 & 0.34
& 16 &  14 &   9 & 23 & 1.007 \\
$^{154}$Sm & 3.25 & 3.28 & 13.4 & 13.5 & 17.6 & 18.6 & 8.4 & 13.7 & 0.20
& 16 &   6 &   7 & 17 & 0.484 \\

$^{154}$Gd & 3.02 & 3.20 &  5.5 &  7.4 &  8.1 &  5.1 & 5.4  & 3.4 & 0.31
& 26 & 26  &   7 & 32 & 1.849 \\
$^{156}$Gd & 3.24 & 3.27 & 11.8 & 11.6 & 13.0 & 14.4 & 7.5  & 10.5 & 0.22
& 26 & 12  &   16 & 34 & 0.605 \\
$^{158}$Gd & 3.29 & 3.30 & 15.0 & 15.4 & 14.9 & 15.1 & 8.3 & 10.6 & 0.36
& 12 &  6  &   6 & 14 & 0.224 \\
$^{160}$Gd & 3.30 & 3.31 & 17.6 & 17.8 & 13.1 & 13.1 & 8.6 & 8.8 & 0.44
& 16 &  4  &   8 & 17 & 0.169 \\
$^{162}$Gd & 3.29 & 3.31 & 19.8 & 19.8 & 12.0 & 12.0 & 9.5 & 8.1 & 0.30
& 14 &  0  &   4 & 10 & 0.082 \\

$^{158}$Dy & 3.21 & 3.25 & 10.0 & 10.1 &  9.6 & 10.0 & 6.7  & 7.1 & 0.25 &
28 &  8 &  8 & 25 & 0.495 \\
$^{160}$Dy & 3.27 & 3.28 & 14.7 & 14.5 & 11.1 & 11.8 & 9.2 & 8.2 & 0.18 &
28 & 4 & 23 & 38 & 0.522 \\
$^{162}$Dy & 3.29 & 3.31 & 17.3 & 17.1 & 11.0 & 10.8 & 8.3 & 7.2 & 0.44 &
18 & 14 & 14 & 29 & 0.312 \\
$^{164}$Dy & 3.30 & 3.31 & 22.6 & 22.2 & 10.4 & 10.2 & 13.1 & 6.8 & 0.14 &
20 & 0 & 10 & 19 & 0.188 \\
$^{166}$Dy & 3.31 & 3.31 & 15.0 & 15.1 & 11.2 & 11.2 & 7.9 & 7.5 & 0.52 &
6 & 2 & 5 & 8 & 0.037 \\

$^{160}$Er & 3.10 & 3.20 &  7.1 &  7.9 &  6.8 & 3.9 & 5.6 &  2.5 & 0.30
& 26 & 2 & 5 & 18 & 1.790\\
$^{162}$Er & 3.23 & 3.26 & 10.7  & 11.1 &  8.8 & 9.9 & 7.1 & 7.0 & 0.26
&20 & 4 & 12 & 23 & 0.588 \\
$^{164}$Er & 3.28 & 3.27 & 13.6 & 12.9 &  9.4 & 9.6 & 9.3 & 6.6 & 0.14
&22 & 10 & 18 & 33 & 0.827 \\
$^{166}$Er & 3.29 & 3.30 & 18.1 & 17.8 &  9.8 & 9.6 & 9.7 &  6.5 & 0.23
& 16 & 10 & 14 & 26 & 0.306 \\
$^{168}$Er & 3.31 & 3.32 & 15.3 & 15.5 & 10.3 & 10.2 & 8.1 & 6.7 & 0.59
& 18 & 6 & 8 & 19 & 0.176 \\
$^{170}$Er & 3.31 & 3.29 & 11.3 &  9.9 & 11.9 & 13.1 & 6.5 & 9.0 & 0.17
&24 & 10 & 19 & 35 & 0.864 \\

\hline
\end{tabular}
\end{table}

\begin{table}
\setcounter{table}{1} \caption{(continued)}

\bigskip

\begin{tabular}{ l r r r  r r r r r r r r r r r}
\hline nucleus & $R_{4/2}$ & $R_{4/2}$ & $R_{0/2}$& $R_{0/2}$
&$R_{2/2}$ & $R_{2/2}$ & $\beta_0$ & $C$   & $a$ &
$L_g$ & $L_\beta$ & $L_\gamma$ & $n$ & $\sigma$ \\
        & exp &  th  & exp & th  & exp & th &  &  &  &  &  &  & &  \\

\hline

$^{164}$Yb & 3.13 & 3.21 & 7.9 & 7.4 & 7.0 & 7.3 & 5.4 &  5.2 & 0.32 &
18 & 0 & 5 & 13 & 0.471\\
$^{166}$Yb & 3.23& 3.25 & 10.2 & 10.1&  9.1 & 9.5 & 6.8 &  6.7 & 0.23 &
24 & 10 & 13 & 29 & 0.688 \\
$^{168}$Yb & 3.27 & 3.27 & 13.2& 11.9  & 11.2 &11.4 & 7.9 & 8.0 & 0.19 &
34 & 4 & 7 & 25 & 0.768 \\
$^{170}$Yb & 3.29 & 3.31& 12.7 & 13.8 & 13.6 &13.5 & 7.8 & 9.2 & 0.49 &
20 & 10 & 17 & 31 & 0.509 \\
$^{172}$Yb & 3.31 & 3.29 & 13.2 & 12.4& 18.6 & 19.1 & 7.8  & 13.7 & 0.17 &
16 & 12 & 5 & 18 & 0.851 \\
$^{174}$Yb & 3.31 & 3.31 & 19.4 & 19.3& 21.4 & 21.5 & 17.2 & 15.1 & 0.05 &
20 & 4 & 5 & 16 & 0.170 \\
$^{176}$Yb & 3.31 & 3.32& 13.9 & 14.0   & 15.4 & 15.3 & 8.1 & 10.4 & 0.57 &
20 & 2 & 5 & 15 & 0.254 \\
$^{178}$Yb & 3.31 & 3.29& 15.7 & 15.6& 14.5 & 14.5 & 10.2 & 10.3 & 0.15 &
6 & 4 & 2 & 6 & 0.045 \\

$^{168}$Hf & 3.11 & 3.22 & 7.6 & 7.5 & 7.1 & 7.4 & 5.4 & 5.3 & 0.32 &
22 & 4 & 4 & 16 & 0.438\\
$^{170}$Hf & 3.19 & 3.23 & 8.7 & 8.6 & 9.5 & 8.1 & 6.0 & 5.7 & 0.27 &
34 & 4 & 4 & 22& 0.964\\
$^{172}$Hf & 3.25 & 3.26 &  9.2 & 10.1 & 11.3 & 11.7 & 6.7  & 8.5 & 0.24 &
38 & 4 & 6 & 26 & 0.444 \\
$^{174}$Hf & 3.27 & 3.26 &  9.1 & 9.0 & 13.5 & 14.1 & 6.1 & 10.4 & 0.26 &
26 & 26 & 5 & 30 & 0.484 \\
$^{176}$Hf & 3.28 & 3.28 & 13.0 & 12.3 & 15.2 & 16.1 & 7.9 & 11.7 & 0.18 &
18 & 10 & 8 & 21 & 0.622 \\
$^{178}$Hf & 3.29 & 3.28 & 12.9 & 12.7 & 12.6 & 12.9 & 8.5 & 9.2 & 0.16 &
18 & 6 & 6 & 17 & 0.298 \\
$^{180}$Hf & 3.31 & 3.31 & 11.8 & 12.2 & 12.9 & 12.9 & 7.7 & 8.7 & 0.61 &
12 & 4 & 5 & 12 & 0.199 \\

$^{176}$W & 3.22 & 3.24 & 7.8 & 8.2 & 9.6 & 10.2 & 5.8 & 7.4 & 0.29 &
22& 12 & 5 & 21 & 0.578\\
$^{178}$W  & 3.24 & 3.25&  9.4 & 9.5 & 10.5 & 10.5& 6.4  & 7.6 & 0.26 &
 18 & 10 & 2 & 15 & 0.177 \\
$^{180}$W  & 3.26 & 3.27& 14.6& 12.7 & 10.8 & 11.4 &7.9 &  8.0 & 0.24 &
24 & 0 & 7 & 18 & 0.838 \\
$^{182}$W  & 3.29 & 3.28 & 11.3& 11.6 & 12.2 & 12.4 &8.3 &  8.6 & 0.13 &
18 & 4 & 6 & 16 & 0.282 \\
$^{184}$W  & 3.27 & 3.27 &  9.0 & 8.9&  8.1 & 8.1 & 6.6 &  5.5 & 0.15 &
 10 & 4 & 6 & 12 & 0.094 \\
$^{186}$W  & 3.23 & 3.24&  7.2 & 7.2 &  6.0 & 6.2 &5.3 &  4.2 & 0.22 &
14 & 4 & 6 & 14 & 0.133 \\

$^{178}$Os & 3.02 & 3.19 & 4.9 & 5.8 & 6.6 & 7.1 & 4.8 & 5.1 & 0.39 &
16& 6 & 5 & 15 & 0.724\\
$^{180}$Os & 3.09 & 3.23 & 5.6 & 7.1 & 6.6 & 7.0 & 5.1 & 4.9 & 0.27 &
10& 6 & 7 & 14 & 1.122\\
$^{184}$Os & 3.20 & 3.24 &  8.7 & 9.4 &  7.9 & 8.5 &6.3 & 6.0 & 0.28 &
22 & 0 & 6 & 16 & 0.600 \\
$^{186}$Os & 3.17 & 3.22 &  7.7 & 7.7 &  5.6 & 6.1 &5.6  &  4.2 & 0.27 &
14 & 10 & 13 & 24 & 0.200 \\
$^{188}$Os & 3.08 & 3.20 &  7.0 & 7.3 &  4.1 & 4.3 &5.3 &  2.8 & 0.34 &
12 & 2 & 7 & 13 & 0.213 \\

\hline
\end{tabular}
\end{table}

\begin{table}
\setcounter{table}{1} \caption{(continued)}

\bigskip

\begin{tabular}{ l r r r  r r r r r r r r r r r}
\hline nucleus & $R_{4/2}$ & $R_{4/2}$ & $R_{0/2}$& $R_{0/2}$
&$R_{2/2}$ & $R_{2/2}$ & $\beta_0$ & $C$   & $a$ &
$L_g$ & $L_\beta$ & $L_\gamma$ & $n$ & $\sigma$ \\
        & exp &  th  & exp & th  & exp & th &  &  &  &  &  &  & &  \\

\hline

$^{228}$Ra & 3.21 & 3.27 & 11.3 & 11.3 & 13.3 & 13.0 & 7.3 & 9.5 & 0.24
& 22 & 4 & 3 & 15 & 0.387 \\

$^{228}$Th & 3.24 & 3.28& 14.4 & 13.9 & 16.8 & 17.0 & 8.4 & 12.4 & 0.25
& 18 & 2 & 5 & 14 & 0.514 \\
$^{230}$Th & 3.27 & 3.28 & 11.9 & 11.7& 14.7 &14.8 & 7.6  & 10.7 & 0.19
& 24 & 4 & 4 & 17 & 0.276 \\
$^{232}$Th & 3.28 & 3.29 & 14.8 & 14.7 & 15.9 & 16.5 & 9.6 & 11.8 & 0.15
& 30 & 20 & 12 & 36 & 0.321 \\

$^{232}$U  & 3.29 & 3.29 & 14.5 & 14.4 & 18.2 & 18.3 & 9.5 & 13.1 & 0.14
&20 & 10 & 4 & 18 & 0.158 \\
$^{234}$U  & 3.30 & 3.30 & 18.6 & 18.7 & 21.3 & 21.7 & 12.1 & 15.5 & 0.12
& 28 & 8 & 7 & 24 & 0.219 \\
$^{236}$U  & 3.30 & 3.30 & 20.3& 20.4 & 21.2 & 21.2 & 13.8 & 15.0 & 0.10
& 30 & 4 & 5 & 21 & 0.266 \\
$^{238}$U  & 3.30 & 3.31 & 20.6 & 20.9 & 23.6 & 24.7 & 13.9 & 17.6 & 0.10
& 30 & 4 & 27 & 43 & 0.716 \\

$^{238}$Pu & 3.31 & 3.32 & 21.4 & 21.7 & 23.3 & 23.4 & 9.8 & 15.9 &  0.50
&26 & 2 & 4 & 17 & 0.857 \\
$^{240}$Pu & 3.31 & 3.31 & 20.1 & 19.1 & 26.6 & 27.1 & 13.0 & 19.4 & 0.09
& 26 & 4 & 4 & 18 & 0.539 \\
$^{242}$Pu & 3.31 & 3.32 & 21.5& 21.6 & 24.7 & 24.8 & 9.9 & 17.0 & 0.46
& 26 & 2 & 2 & 15 & 0.926 \\

$^{248}$Cm & 3.31 & 3.31& 25.0 & 25.1& 24.2 & 24.2& 16.4 & 17.0 & 0.09
&28 & 4 & 2 & 17 & 0.105 \\

$^{250}$Cf & 3.32 & 3.32& 27.0 & 27.0 & 24.2 & 24.2& 12.8 & 16.7 & 0.20
&8 & 2 & 4 & 8 & 0.018 \\

\hline
\end{tabular}
\end{table}

\begin{table}

\caption{Theoretical predictions of the $\gamma$-unstable Morse
potential with $\beta_0=4.22$ and $a=0.38$ compared to experimental
data for $^{154}$Dy \cite{NDS}, and theoretical predictions of the
exactly separable Morse (ES-M) with $\beta_0=9.6$, $C=11.8$,
$a=0.15$ compared to experimental data for $^{232}$Th \cite{NDS}.
All states are normalized to the $2_1^+$ state. See Section V for
further discussion.}

\begin{tabular}{ r r r r r r r r r | r r r r r }
\hline
   & $^{154}$Dy & $^{154}$Dy & $^{232}$Th & $^{232}$Th & $^{154}$Dy & $^{154}$Dy & $^{232}$Th & $^{232}$Th &  &
     $^{154}$Dy & $^{154}$Dy & $^{232}$Th & $^{232}$Th \\
$L$& gsb & gsb & gsb & gsb & $\beta_1$ & $\beta_1$ & $\beta_1$ & $\beta_1$ & $L$ & $\gamma_1$ & $\gamma_1$ &
$\gamma_1$ & $\gamma_1$ \\
   &    exp &  th    & exp    &  th    &  exp &  th    & exp    &  th    &   &    exp &  th   &    exp &  th \\
\hline
0 &  0.00 &  0.00 &  0.00 &  0.00 & 1.98 & 2.66 & 14.79 & 14.72 &  2 &  3.07 & 2.28 & 15.91 & 16.48 \\
2 &  1.00 &  1.00 &  1.00 &  1.00 & 2.71 & 3.35 & 15.68 & 15.63 &  3 &  3.99 & 3.71 & 16.80 & 17.32 \\
4 &  2.23 &  2.28 &  3.28 &  3.29 & 3.74 & 4.24 & 17.68 & 17.72 &  4 &  4.31 & 3.71 & 18.03 & 18.42 \\
6 &  3.66 &  3.71 &  6.75 &  6.76 & 4.96 & 5.25 & 20.72 & 20.90 &  5 &  5.20 & 5.25 & 19.45 & 19.79 \\
8 &  5.22 &  5.25 & 11.28 & 11.29 & 6.47 & 6.37 & 24.75 & 25.06 &  6 &  5.64 & 5.25 & 21.27 & 21.40 \\
10&  6.89 &  6.86 & 16.75 & 16.73 & 8.24 & 7.58 & 29.76 & 30.08 &  7 &  6.53 & 6.86 & 23.21 & 23.26 \\
12&  8.65 &  8.56 & 23.03 & 22.97 &      &      & 35.55 & 35.86 &  8 &       &      & 25.50 & 25.35 \\
14& 10.49 & 10.37 & 30.04 & 29.89 &      &      & 42.14 & 42.32 &  9 &       &      & 27.75 & 27.65 \\
16& 12.47 & 12.30 & 37.65 & 37.43 &      &      & 49.44 & 49.38 & 10 &       &      & 30.62 & 30.17 \\
18& 14.55 & 14.36 & 45.84 & 45.52 &      &      & 57.36 & 57.03 & 11 &       &      & 33.22 & 32.88 \\
20& 16.71 & 16.58 & 54.52 & 54.16 &      &      & 65.81 & 65.23 & 12 &       &      & 36.48 & 35.78 \\
22& 18.98 & 18.95 & 63.69 & 63.31 &      &      &       &       &    &       &      &       &       \\
24& 21.40 & 21.48 & 73.32 & 72.99 &      &      &       &       &    &       &      &       &       \\
26& 23.99 & 24.19 & 83.38 & 83.20 &      &      &       &       &    &       &      &       &       \\
28&       &       & 93.82 & 93.94 &      &      &       &       &    &       &      &       &       \\
30&       &       &104.56 &105.24 &      &      &       &       &    &       &      &       &       \\
\hline
\end{tabular}
\end{table}

\end{document}